# Magnetic doping of a thiolated-gold superatom

De-en Jiang*

*Chemical Sciences Division, Oak Ridge National Laboratory, Oak Ridge, TN, 37831*

Robert L. Whetten

*School of Chemistry & Biochemistry, Georgia Institute of Technology, Atlanta, Georgia 30332*

**Abstract**

The $Au_{25}(SR)_{18}^-$ cluster is a new member in the superatom family which features a centered icosahedral shell ($Au_{13}$) protected by six $RS(AuSR)_2$ motifs (RS− being an alkylthiolate group). Here we show that this superatom can be magnetically doped by replacing the center Au atom with Cr, Mn, or Fe. We find that Cr and Mn-doped clusters have an optimized magnetic moment of 5 Bohr magnetons while the Fe-doped cluster has an optimized magnetic moment of 3 Bohr magnetons. Although the dopant atom's local magnetic moment makes a major contribution to the total moment, the icosahedral $Au_{12}$ shell is also found to be significantly magnetized. Our work here provides a new scenario of magnetic doping of a metal-cluster superatom which is protected by ligands and made by wet chemistry.

*To whom correspondence should be addressed. E-mail: jiangd@ornl.gov. Phone: (865)574-5199. Fax: (865) 576-5235.

## I. INTRODUCTION

Metal clusters generated in the gas phase often have multi-shell structures and special stability at certain compositions.[1] These so-called magic numbers can be explained by the electronic shell model,[2] so the whole cluster can be viewed as a superatom.[3,4] The multi-shell structures are subject to doping by foreign atoms, and how the dopant's orbitals interact with the rest of the cluster is an interesting question. Numerous experimental and computational efforts have targeted this problem.[5-15]

Recently, the superatom concept has been successfully extended to explain the special stability of certain-sized, ligated gold clusters prepared by wet chemistry.[16,17] This new development is exciting in that the wet-chemistry synthesis allows much wider chemical tuning and hence enjoys much broader applications[18] than the gas-phase generation of metal clusters. One such ligated gold magic cluster is $Au_{25}(SR)_{18}^-$ which features a centered icosahedral shell ($Au_{13}$) protected by 6 RS(AuSR)$_2$ motifs (RS− being an alkylthiolate group).[19,20] This cluster has an electron count of 8 which occupies the 1S and 1P levels of the superatom, hence very stable.[17]

The $Au_{25}(SR)_{18}^-$ cluster can be doped by replacing the center Au atom by a foreign atom (Fig. 1). By tuning the cluster charge, one can maintain the 8-electron count and therefore the geometry and electronic structure of the original cluster. This idea has been motivated by an experimental study[21] and pursued computationally by several groups independently.[22-24] It has been predicted that many transition metals (such as Pd, Pt, and Cu)[22-24] and main-group elements (such as Be and Al)[22] can be good dopants of the $Au_{25}(SR)_{18}^-$ superatom.

The doping studies of the $Au_{25}(SR)_{18}^-$ superatom discussed above, however, deal with the nonmagnetic doping only; namely, the doped cluster is nonmagnetic and the dopant atoms all have a closed electronic shell in the cluster. Here one asks whether the $Au_{25}(SR)_{18}^-$ superatom can be doped magnetically. Since some atoms (such as Cr and Mn) have a stable $d^5$ configuration, one obvious idea is to use such atoms to replace the center Au atom in $Au_{25}(SR)_{18}^-$ and then maintain the dopant's $d^5$ configuration and the cluster's 8-electron count by tuning the cluster charge. In this paper, we pursue this idea of magnetic doping by electron accounting supported by first principles density functional theory. We will show that the $Au_{25}(SR)_{18}^-$ superatom can indeed be magnetically doped.

II. COMPUTATIONAL METHODS

We employed both the Vienna Ab Initio Simulation Package (VASP)[25,26] and Turbomole[27] V5.10 to perform DFT calculations. VASP employs periodic boundary conditions and planewave bases. We first put the cluster in a cubic box (25×25×25 Å$^3$) and used VASP to simultaneously optimize the structure *and magnetic moment* for the cluster. After a converged structure and magnetic moment were obtained, we then used Turbomole to re-optimize the cluster's structure at the VASP-optimized magnetic moment. We also used Turbomole for the analysis of orbitals and energetics. Therefore, the structure, energetics, and orbital levels reported in this work are from Turbomole, while the optimized magnetic moments are from VASP.

The Perdew-Burke-Erzonhoff (PBE) form of the generalized-gradient approximation (GGA) was chosen for electron exchange and correlation[28] for both VASP and Turbomole calculations. For VASP, the electron-core interaction was described by the projector-augmented wave (PAW) method within the frozen-core approximation,[29,30] kinetic energy cutoff was set at 450 eV which was found to be sufficient to converge the energy, and force convergence criterion for geometry optimization was set at 0.05 eV/Å. For Turbomole, default orbital and auxiliary basis sets [def2-SV(P)] were used for all atoms for structural optimization (force convergence criterion at $1.0 \times 10^{-3}$ a.u.), and an effective core potential which includes scalar relativistic corrections was used for Au. Accuracy of def2-SV(P) basis sets is similar to that of 6-31G*.

III. RESULTS AND DISCUSSION

Doping the $Au_{25}(SR)_{18}^-$ superatom is achieved by replacing the center Au atom with a dopant atom M (Fig. 1). The nonmagnetic doping has been demonstrated from first principles DFT studies by several groups.[22-24] A simple rule to maintain the 8-electron-count of the $Au_{25}(SR)_{18}^-$ superatom is x = q + 2, where q is the charge of the cluster and x is the number of valence electrons of the dopant atom.[22] This rule applies to the nonmagnetic doping. If one considers the half-filled $d$ shell (that is, $d^5$) to be a typical case of the magnetic doping, then one adds five more electrons to the valence electron requirement and arrives at x = q + 7. So for q = −1, 0, and +1, x should be 6, 7, and 8, which corresponds to Cr, Mn, and Fe, respectively (here we consider only 3$d$ metals and also exclude |q|>1 scenarios). Hence, we obtain three candidates for the magnetic superatom: $Cr@Au_{24}(SR)_{18}^-$, $Mn@Au_{24}(SR)_{18}$, and $Fe@Au_{24}(SR)_{18}^+$.

One can also understand how we arrived at the candidates above from the electronic configuration for the dopant atom ($d^5s^1$, $d^5s^2$, and $d^6s^2$ for Cr, Mn, and Fe, respectively). The neutral $Au_{24}(SR)_{18}$ frame contributes 6 (= 24 – 18) electrons[16] to the 8-electron count and hence is 2 electrons short. For $Cr@Au_{24}(SR)_{18}^-$, Cr donates its one 4s electron and the negative charge contributes another electron to fulfill the 8-electron count of the superatom, and the remaining $d^5$ configuration for the Cr atom leads to the desired 5 $\mu_B$ magnetic moment of the cluster. For Mn@Au$_{24}$(SR)$_{18}$, the case is similar, and the difference is that the 2-elecron deficiency of the 8-electron count is provided all by Mn's two 4s electrons. For Fe@Au$_{24}$(SR)$_{18}^+$, one needs to take away one *3d* electron in order to achieve the $d^5$ configuration.

To verify the preceding reasoning, we optimized the magnetic moments and structures for the three candidates and found that both Cr@Au$_{24}$(SR)$_{18}^-$ and Mn@Au$_{24}$(SR)$_{18}$ have an optimized magnetic moment of 5 $\mu_B$ (Table I), indicating that the $d^5$ configuration of the dopant atom is well preserved, as we have designed. The other spin states (with total magnetic moments of 1, 3, and 7 $\mu_B$) are rather higher in energy (Table II). The icosahedral Au$_{12}$ shell in both clusters is also well maintained: with a tolerance of 0.25 Å, both clusters' shells show I$_h$ symmetry; with a tolerance of 0.1 Å, Cr@Au$_{24}$(SR)$_{18}^-$'s Au$_{12}$ shell shows D$_{2h}$ symmetry while that of Mn@Au$_{24}$(SR)$_{18}$ shows T$_h$ symmetry. Fe@Au$_{24}$(SR)$_{18}^+$, however, shows an optimized magnetic moment of 3 $\mu_B$, and we found that the spin sextet state (that is, total magnetic moment at 5 $\mu_B$) is 0.16 eV higher in energy. The Au$_{12}$ shell of the Fe@Au$_{24}$(SR)$_{18}^+$ cluster at the optimized magnetic moment of 3 $\mu_B$ was found to deform significantly: the shell shows C$_i$ symmetry with a tolerance of 0.1 Å and S$_6$ symmetry with a tolerance of 0.25 Å. This deformation also leads to significantly varying M-Au distances, evidenced by the larger standard deviation of the M-Au distance (Table I).

The $d^5$ configuration we have intended to preserve by magnetic doping of the Au$_{25}$(SR)$_{18}^-$ superatom is achieved for Cr and Mn. Certainly, the orbital levels of the doped cluster should explain the 5 $\mu_B$ magnetic moment. We use Cr as an example to show orbital levels. Fig. 2 displays the spin-up and spin-down orbital levels of the Cr@Au$_{24}$(SR)$_{18}^-$ cluster near the Fermi level. By inspecting these orbitals, we found that the three highest occupied spin orbitals for both up and down spins can be described as the 1P levels of the superatom and the next five occupied spin-up orbitals show major character of the d states of the center Cr atom. One can also see that the frontier superatomic levels and the Cr d-dominated levels for the up spin are well separated, which leads to the cluster's 5 $\mu_B$ magnetic moment. Moreover, the 1P levels are slightly split due to the non-ideal icosahedral Au$_{12}$ shell. The five Cr d-dominated states are split into the familiar t$_{2g}$ and e$_g$ orbitals.

We now examine what typical frontier spin up orbitals for the Cr@Au$_{24}$(SR)$_{18}^-$ cluster look like. Fig. 3a shows the highest occupied spin up orbital and one can see that it mainly locates at the Au$_{12}$ shell, with some contribution from the sulfur atoms. This is consistent with this orbital's 1P character. Fig. 3b shows the highest occupied spin up orbital among the five Cr d-dominated levels described in Fig. 2. One can see that this orbital is indeed centered on the Cr atom and displays d$_z^2$ character. It also has contributions from the Au$_{24}$(SR)$_{18}$ framework. Fig. 4 shows the spin magnetization density for the Cr@Au$_{24}$(SR)$_{18}^-$ cluster and one can see that the isosurface has a spherical shape around the center Cr atom, indicating that the center atom is mainly responsible for the cluster's magnetic moment. The computed local magnetic moment at the Cr center is 3.53 $\mu_B$ (Table I), indicating that there are also contributions from the Au$_{24}$(SR)$_{18}$ frame to the total 5 $\mu_B$ moment. We found that this is indeed the case and the Au$_{12}$ shell is magnetized with an average local magnetic moment of ~0.1 $\mu_B$ on the shell gold atoms.

This local magnetic moment is caused by spin magnetization, namely, the nonzero spin density around a shell gold atom. This can be further understood by examining again one of the five Cr d-dominated spin up levels (Fig. 3b). This orbital can be viewed as contributing one unpaired electron (that is, one Bohr magneton) to the total spin, and one can see that it does have significant distribution at some shell gold atoms.

We next examine the thermodynamic driving force for magnetic doping of the $Au_{25}(SR)_{18}^-$ superatom. We computed the interaction energy ($E_{Int}$) between the dopant atom and the $Au_{24}(SR)_{18}^q$ frame from the following equation:

$$E_{Int} = E(M) + E[Au_{24}(SR)_{18}^q] - E[M@Au_{24}(SR)_{18}^q] \qquad (1),$$

where $E(M)$, $E[Au_{24}(SR)_{18}^q]$, and $E[M@Au_{24}(SR)_{18}^q]$ are the energies of an isolated dopant atom, the $Au_{24}(SR)_{18}^q$ frame (whose atomic positions are taken from $M@Au_{24}(SR)_{18}^q$ and not relaxed), and the $M@Au_{24}(SR)_{18}^q$ cluster, respectively. Table I shows the interaction energies for the three dopants. One can see that the interaction between the dopant atom and the frame is rather strong for all three dopants. The magnitude of the interaction energy for the three magnetic dopants (between 6 and 7 eV) can be compared with that for the nonmagnetic dopants. For example, we found that Ni, Cu, and Zn have an interaction energy of 8.5, 5.8, and 4.3 eV, respectively.[22] In addition, Au itself has an interaction energy of 4.0 eV,[22] so replacing the center Au atom by Cr actually gains over 2 eV, indicating that doping by magnetic dopants such as Cr is thermodynamically very favorable.

The orbital-level distribution of the $Mn@Au_{24}(SR)_{18}$ cluster is similar to that of the $Cr@Au_{24}(SR)_{18}^-$ cluster, but the separation between the dopant's *d* states with the rest is not as clear as in $Cr@Au_{24}(SR)_{18}^-$. In other words, the interaction between the dopant's 3$d$ states and the Au$_{24}$(SR)$_{18}$ frame is stronger in Mn@Au$_{24}$(SR)$_{18}$. This interaction becomes even stronger in Fe@Au$_{24}$(SR)$_{18}^{+}$, which leads to decreased magnetic moment than the ideal 5 μ$_B$ of the d$^5$ configuration. Indeed, we found that the second highest occupied spin-down orbital of Fe@Au$_{24}$(SR)$_{18}^{+}$ has a major contribution from an Fe 3$d$ orbital. In other words, the Fe dopant's d$^5$ configuration consists of one spin down electron and four spin up d electrons, resulting in a magnetic moment of 3μ$_B$ for the cluster (Table I). One also notes from Table I that the local magnetic moment on Fe is greater than 3μ$_B$, which means that some atoms in the Au$_{24}$(SR)$_{18}$ frame are antiferromagnetically coupled to the Fe atom. In need, we found that some Au atoms in the icosahedral shell have negative spin densities.

Due to its rigid structure, the icosahedral Au$_{12}$ shell can be a good protecting layer for the center atom, and the shell itself can be further protected by an outside layer of ligands. Experimentally, a halogen and phosphine ligand-protected Pd@Au$_{12}$ cluster has been synthesized,[31] which is encouraging for realizing M@Au$_{24}$(SR)$_{18}^{q}$ clusters. In fact, Murray and coworkers[21] had been experimenting the doping of the Au$_{25}$(SR)$_{18}^{-}$ superatom by Pd even before several computational studies of doping the Au$_{25}$(SR)$_{18}^{-}$ superatom were started.[22-24]

Recently, Jin and coworkers showed that the closed-shell Au$_{25}$(SR)$_{18}^{-}$ superatom can be reversibly oxidized to the neutral form which is a doublet, thereby displaying switchable magnetism between nonmagnetic (anionic) and paramagnetic (neutral) states.[32] The magnetically doped clusters examined in the present work should have richer magnetic properties and redox chemistry due to their high-spin center atoms. This is a potentially interesting topic if the predicted magnetic clusters here can be realized.

We note that the magnetic clusters predicted here are supposed to be paramagnetic[32] and therefore distinct from single-molecule magnets which are polynuclear metal complexes comprising several transition-metals ions bridged by oxygen at the core, such as $Mn_{12}O_{12}$, and exhibiting superparamagnetic-like properties.[33]

Magnetic doping of bare metal clusters has been demonstrated both experimentally and computationally.[5-15] The dopant's magnetic moment can be either fully quenched or partly reduced or fully preserved or enhanced. Lievens and coworkers examined transition-metal doped Au and Ag clusters with photofragmentation and mass spectrometry.[7-9] They found that the electron-shell model can explain the magic numbers in the observed doped clusters. For example, in their Cr, Mn, and Fe-doped Au clusters,[7,8] the fragmentation spectra indicated that the dopant contributes their two 4$s$ electrons to close the shell, which suggests that these clusters should be magnetic due to the unpaired 3$d$ electrons of the dopant. Similar discoveries were also made by Wang and coworkers[10] in their photoelectron spectroscopic study of Ti, V, and Cr-doped $Au_6^-$. In the doped Ag clusters, Lievens and coworkers[9] found special stability for $Ag_{10}Co^+$, which suggests that the magnetic moment on Co is quenched and both its 3$d$ and 4$s$ electrons contribute to the 18 electron count.

In the computational studies of doping bare metal clusters, Sun et al.[5] examined 3d and 4d dopants in a $Cu_{12}$ shell in both $I_h$ and $O_h$ symmetries computationally by using the DFT method and found that Cr's magnetic moment is fully quenched because its six valence electrons all contribute to the 18 shell-closing electron count, while Fe and Mn's magnetic moments are reduced to 1.32 and 1.70 $\mu_B$, respectively, in the icosahedral $Cu_{12}$ shell. For clusters of fewer atoms, Li et al.[10] found that Ti, V, and Cr maintain large magnetic moments (2, 3, and 4 $\mu_B$, respectively) in the center of the planar $Au_6$ ring for the $Au_6M^-$ cluster. Janssens et al.[8,9] and Torrres et al.[11] predicted that 3d dopants such as Cr, Mn, and Fe all have significant magnetic moments in $Ag_5M^+$ and $Au_5M^+$. To explain the observed abundance peaks in the photofragmentation peaks of doped clusters,[7] Torres et al.[11] examined magnetic properties of $Au_nM^+$ clusters (for n < 10) and concluded that the magnetic and geometric configurations are strongly correlated. Pradhan et al.[14] examined Sc, Ti, and V doped $Na_n$ (n=4, 5, 6) clusters with DFT and found that the dopant's magnetic moment is enhanced upon its free-atom value, which is attributed to hybridization between the dopant's $d$ states and the alkali metal's sp states. Recently, Wang et al.[15] studied magnetic doping of the cage cluster $Au_{16}^-$ by Fe, Co, and Ni with trapped ion electron diffraction, photoelectron spectroscopy, and DFT, and they found that high spin moments can be retained for the dopants.

The doped clusters discussed in the preceding paragraph are, however, based on the gas-phase experiments of cluster generation. Although the high spin moments predicted in the present work for Cr, Mn, and Fe dopants are not surprising, the importance of magnetically doping the $Au_{25}(SR)_{18}^-$ superatom lies in the fact that $Au_{25}(SR)_{18}^-$ is prepared via wet chemistry and very stable.[19,20,34-38] In fact, single crystals for $Au_{25}(SR)_{18}^-$ have been obtained.[19,20] This implies that one may make air- and thermal-stable $M@Au_{24}(SR)_{18}$ in a very large scale and subject it to sophisticated physical measurements. Recent work by Jin and coworkers[32,38] clearly demonstrates the opportunities along this direction of research. Our work here hints that doping the $Au_{25}(SR)_{18}^-$ superatom by Cr, Mn, and Fe is worth pursuing and may offer some exciting opportunities for magnetism in a ligated metallic cluster.

Further, gold nanoparticles have many biomedical applications, such as sensing, imaging, therapy, and drug delivery.[18,39] With a large magnetic moment and a small size, M@Au$_{24}$(SR)$_{18}$ clusters (M being Cr or Mn) as a paramagnetic molecule may find some special applications.

IV. SUMMARY AND CONCLUSIONS

We have shown that the Au$_{25}$(SR)$_{18}^-$ superatom can be magnetically doped by Cr, Mn, and Fe. Both Cr and Mn-doped clusters have an optimized magnetic moment of 5 $\mu_B$, indicating that the d$^5$ configuration of the dopant atom is well preserved at the cluster center with all electrons spin up. This is confirmed by orbital analysis. The structure of the original Au$_{24}$(SR)$_{18}$ frame is also well maintained for Cr and Mn. So Cr@Au$_{24}$(SR)$_{18}^-$ and Mn@Au$_{24}$(SR)$_{18}$ can be described a magnetic superatom. Fe-doped cluster shows an optimized magnetic moment of 3 $\mu_B$, resulting from stronger interaction between the superatomic levels and the Fe $d$ states, which leads to occupation of a spin-down $d$ state and larger deformation of the Au$_{24}$(SR)$_{18}$ frame. Computed interaction energy between the dopant atom and the Au$_{24}$(SR)$_{18}^q$ frame indicates that magnetic doping by Cr, Mn, and Fe is all thermodynamically favorable. Our work brings about a new aspect of magnetic doping of superatoms by showing that the wet-chemistry prepared Au$_{25}$(SR)$_{18}^-$ superatom can be magnetically doped.

**Added note**: After we submitted this paper, a report on "Designer magnetic superatoms" was published.[40] In it, the authors predicted VCs$_8$ and MnAu$_{24}$(SH)$_{18}$ to be magnetic superatoms. Their MnAu$_{24}$(SH)$_{18}$ cluster has the same structure as our Mn@Au$_{24}$(SR)$_{18}$, except that we used –SCH$_3$ for –SR while they used –SH.

**Acknowledgement**

This work was supported by the Office of Basic Energy Sciences, U.S. Department of Energy under Contract No. DE-AC05-00OR22725 with UT-Battelle, LLC. D.E.J. thanks Prof. Gangli Wang for very useful discussion. This research used resources of the National Energy Research Scientific Computing Center, which is supported by the Office of Science of the U.S. Department of Energy under Contract No. DE-AC02-05CH11231.

Table I. Magnetic dopants (M), the cluster charge (q), optimized total magnetic moment ($\mu_T$, in Bohr magnetons) of the cluster, local magnetic moment on the dopant ($\mu_M$), the interaction energy ($E_{Int}$), average M-Au distance ($R_{M-Au}$), and average Au-Au ($R_{Au-Au}$) distance in the $Au_{12}$ shell for the $M@Au_{24}(SR)_{18}^q$ cluster.

| M | q | $\mu_T$ | $\mu_M$ | $E_{Int}$ (eV)[a] | $R_{M-Au}$ (Å)[b] | $R_{Au-Au}$ (Å)[b] |
|---|---|---|---|---|---|---|
| Cr | −1 | 5 | 3.53 | 6.18 | 2.84 (0.01) | 2.99 (0.10) |
| Mn | 0 | 5 | 3.92 | 6.52 | 2.84 (0.01) | 2.99 (0.11) |
| Fe | +1 | 3 | 3.01 | 6.54 | 2.86 (0.07) | 3.01 (0.21) |

[a]The interaction energy is defined by Eq. 1 in the text. [b]The numbers in parentheses are standard deviation.

Table II. Relative energies (in eV) of different spin states ($\mu_T$, in Bohr magnetons) of Cr@Au$_{24}$(SR)$_{18}^-$ and Mn@Au$_{24}$(SR)$_{18}$.

| $\mu_T$ | 1 | 3 | 5 | 7 |
|---|---|---|---|---|
| Cr@Au$_{24}$(SR)$_{18}^-$ | 0.93 | 0.48 | 0 | 1.10 |
| Mn@Au$_{24}$(SR)$_{18}$ | 0.72 | 0.32 | 0 | 1.03 |

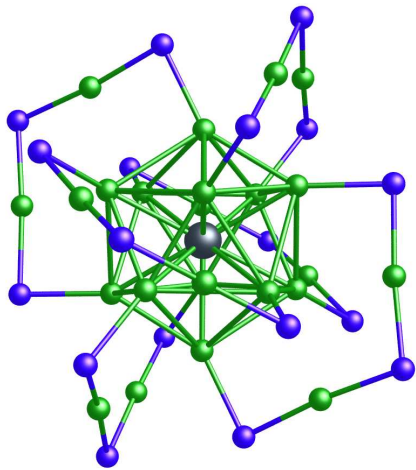

FIG. 1. (Color online) The M@Au$_{24}$(SR)$_{18}^q$ cluster from replacing the center Au atom of Au$_{25}$(SR)$_{18}^-$ with a dopant M. Au, light gray (green); S, dark gray (blue); R−, not shown.

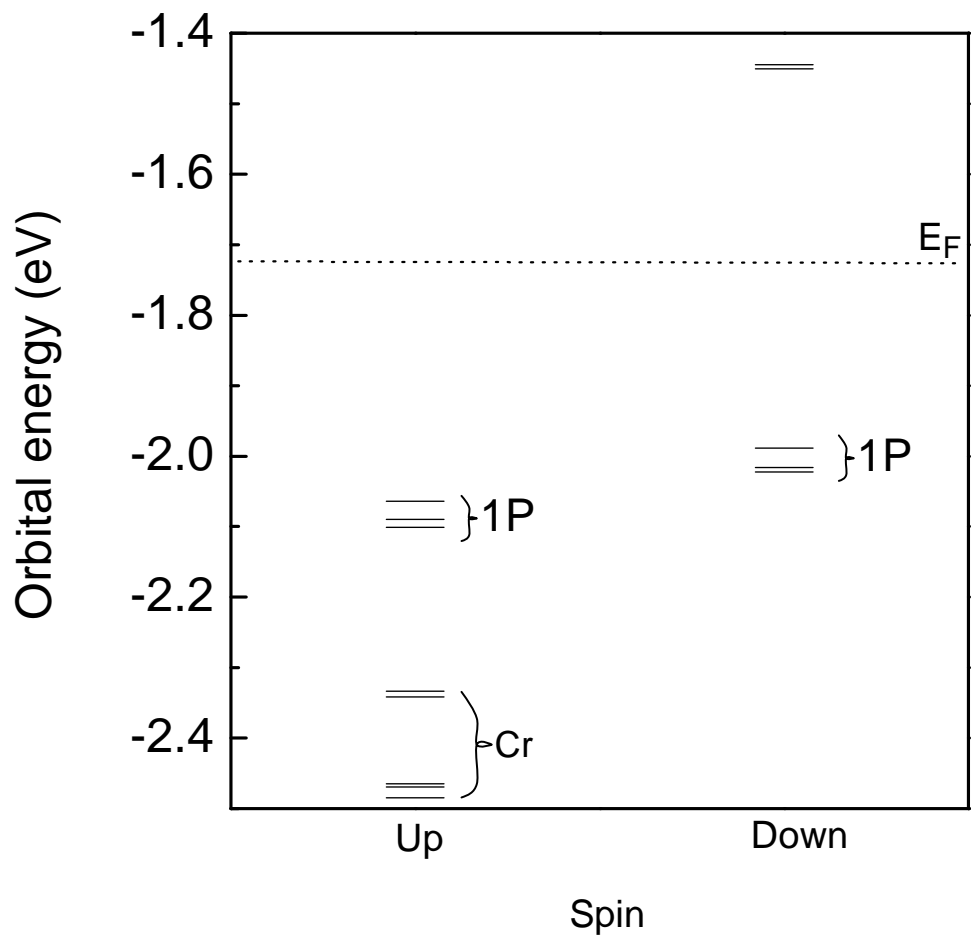

FIG. 2. Frontier spin orbital levels of $Cr@Au_{24}(SCH_3)_{18}^-$.

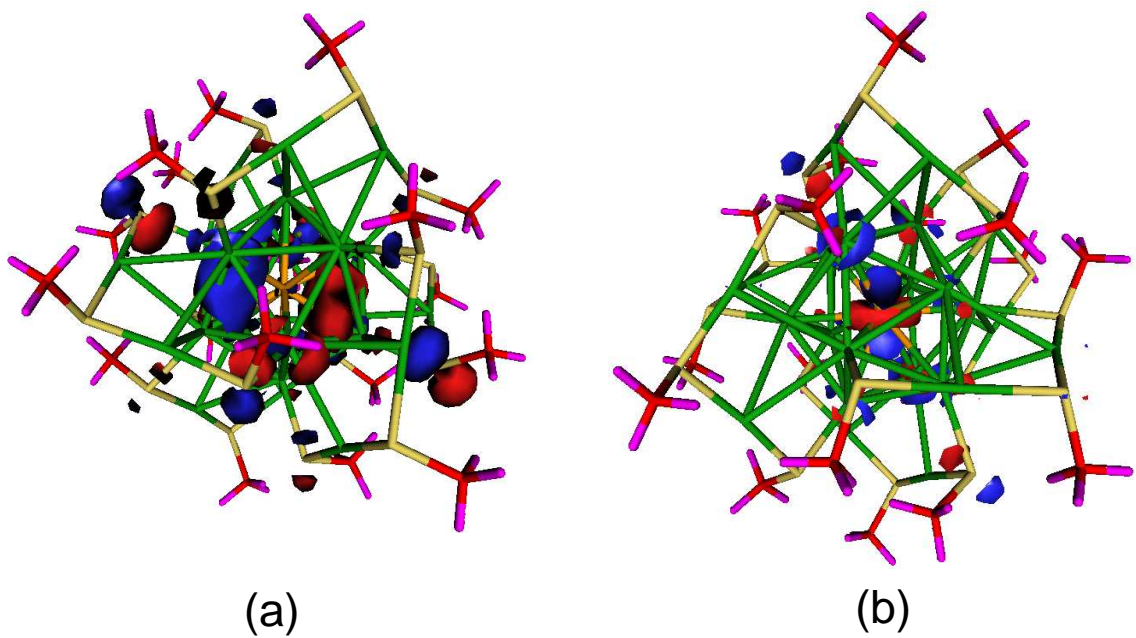

FIG. 3. (Color online) Spin up orbitals of Cr@Au$_{24}$(SCH$_3$)$_{18}^-$: (a) Highest occupied spin up orbital; (b) the highest occupied spin up orbital among the five Cr d-dominated levels (see Fig. 2).

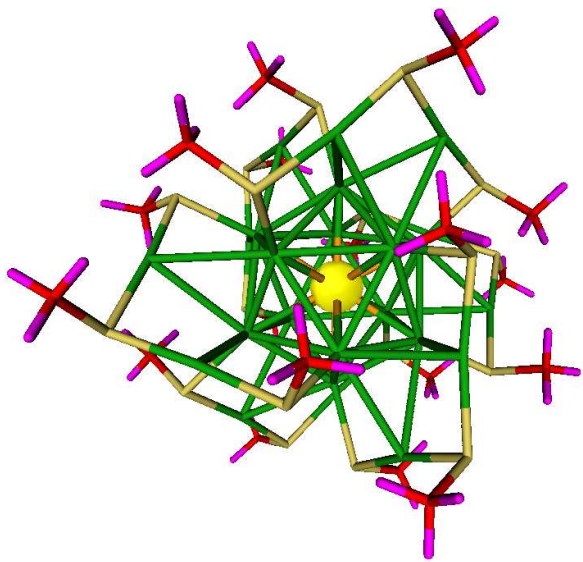

FIG. 4. (Color online). Isosurface plot of spin magnetization density of Cr@Au$_{24}$(SR)$_{18}^{-}$ (the isosurface is the center sphere whose isovalue is at 0.02 a.u.).